\documentclass[%
 preprint,
 amsmath,amssymb,
 aps, physrev,
]{revtex4-2}

\usepackage{graphicx}
\usepackage{dcolumn}
\usepackage{bm}

\begin{document}

\preprint{}

\global\long\def\v#1{\boldsymbol{#1}}%
\global\long\def\T#1{\hat{#1}}%
\global\long\def\d{\text{d}}%
\global\long\def\i{\text{i}}%
\global\long\def\t{\text{t}}%
\global\long\def\nutens{\T{\nu}}%
\global\long\def\mutens{\T{\mu}}%
\global\long\def\v#1{\boldsymbol{#1}}%
\global\long\def\d{\text{d}}%
\global\long\def\i{\text{i}}%
\global\long\def\t{\text{t}}%
\global\long\def\ph{\varphi}%
\global\long\def\tht{\vartheta}%
\global\long\def\balpha{\boldsymbol{\alpha}}%
\global\long\def\btheta{\boldsymbol{\theta}}%
\global\long\def\bJ{\boldsymbol{J}}%
\global\long\def\bGamma{\boldsymbol{\Gamma}}%
\global\long\def\bOmega{\boldsymbol{\Omega}}%
\global\long\def\d{\text{d}}%
\global\long\def\t#1{\text{#1}}%
\global\long\def\m{\text{m}}%
\global\long\def\bm{\text{\textbf{m}}}%
\global\long\def\k{\text{k}}%
\global\long\def\i{\text{i}}%
\global\long\def\c{\text{c}}%
\global\long\def\v#1{\boldsymbol{#1}}%
\global\long\def\difp#1#2{\frac{\partial#1}{\partial#2}}%

\title{Symplectic integration of guiding-center equations\\ in canonical coordinates for general toroidal fields}

\author{Christopher G. Albert$^1$}
\author{Georg S. Graßler$^1$}
\author{Sergei V. Kasilov$^{1,2}$}
\author{Markus Markl$^{1}$}
\author{Jonatan Schatzlmayr$^1$}

\affiliation{%
\,\\$^1$Fusion@ÖAW, Institute of Theoretical and Computational Physics, Graz University of Technology, Graz, Austria\\
%
%
$^2$Institute of Plasma Physics, National Science Center ``Kharkov Institute of Physics and Technology'', Kharkiv, Ukraine
}%

\date{\today}

\begin{abstract}
Symplectic integrators with long-term preservation of integrals of motion are introduced for the guiding-center model of plasma particles in toroidal magnetic fields of general topology. An efficient transformation to canonical coordinates from cylindrical and flux-like coordinates is discussed and applied using one component of the magnetic vector potential as a spatial coordinate. This choice is efficient in both, theoretical and numerical developments and marks a generalization of magnetic flux coordinates. The transformation enables the application of conventional symplectic integration schemes formulated in canonical coordinates, as well as variational integrators on the guiding-center system, without requiring magnetic flux coordinates. Symplectic properties and superior efficiency of the implicit midpoint scheme compared to conventional non-symplectic methods are demonstrated on perturbed tokamak fields with magnetic islands and stochastic regions. The presented results mark a crucial step towards gyrokinetic models that conserve physical invariants.
\end{abstract}

\maketitle


\section{Introduction}

Magnetized plasmas appear in nature as well as science and technology,
notably magnetic confinement fusion. One of the foundations of plasma
theory and computation is the tracing of charged particle orbits in
electromagnetic fields. In the case of strong magnetization, their
gyrofrequency $\omega_{c}$ is large and their gyroradius $\rho_{L}$
is small compared to other scales of frequency and length in the system.
This separation of scales led to the introduction of the guiding-center
system, where the rapid gyromotion $\phi$ appears as an ignorable
variable together with the perpendicular adiabatic invariant $J_{\perp}$
as a constant of motion. Within its range of validity, the guiding-center
ansatz has two main features: it reduces the dimension of the relevant
phase-space and allows numerical treatment at larger time-steps than
the gyration period $2\pi/\omega_{c}$. However, these features come at a price
-- the loss of canonical coordinates.

Why are canonical coordinates important? First, they form the basis for analytical treatment in terms of action-angle variables~\citep{Hazeltine1981-1164,Mahajan1985-3538,Cary1988,Chavanis2007-469}.
Analytical and semi-analytical methods can describe resonant
wave-particle interaction and the emergence of Hamiltonian
chaos~\citep{Chirikov1979-263,Abdullaev2006-S113}. Second, canonical coordinates
underlie an important class of numerical methods:
symplectic integrators~\citep{Hairer2006}. Such integrators
and their generalizations retain a discrete symplectic structure of
phase-space and thereby essential features of continuous Hamiltonian
systems. This includes conservation of energy and momentum within
fixed bounds. In application to alpha particle losses in stellarator
optimization, symplectic integrators outperform conventional
methods by factor 3-5 in computing time and do not require high integration accuracy for orbit classification methods based on the properties of Hamiltonian systems~\citep{Albert2020-109065,Albert2020-815860201,Albert2023-955890301}. 

Symplectic integrators are a subset of more general structure-preserving
methods. Such approaches based on non-canonical coordinates include
variational integrators \citep{Qin2008-35006,Qin2009-42510,Morrison2017-55502,Kraus2017,Burby2017-110703,Ellison2018-52502},
methods based on exact integration in lowest-order fields~\citep{Eder2020-122508,Schatzlmayr2024}
and methods based on slow manifolds \citep{Burby2021-093506,Xiao2021-107981},
line integrals \citep{Brugnano2020-112994} and average vector fields
\citep{Zhu2022-032501,Zhang2024-065101}. Due to the special nature
of the guiding-center system being a degenerate Hamiltonian system
formulated in non-canonical coordinates, various methods
come with specific limitations.
The schemes are either not fully symplectic, e.g. preserve only energy,
are of low order, or require stabilization. Straightforward application
of standard symplectic integrators in canonical coordinates is
of interest to obtain simple, reliable and fully structure-preserving
integration schemes.

For an \emph{explicit} transformation with canonical coordinates given
as functions of non-canonical ones, it is well known~\citep{Hairer2006}
that symplectic integrators can be applied in a straight-forward manner.
This has been leveraged for analytical approximations of fields in
tokamak plasmas~\citep{Khan2012-547,Khan2015-298,Khan2017-40}. Also
for an \emph{implicit} transformation to canonical coordinates it
is possible to formulate symplectic integrators. The original idea
by~\citet{Cary1989} of solving the implicit transformation together
with the implicit time-step of a symplectic integrator has
been picked up recently~\citep{Zhang2014-32504,Zhu2016-1650008,Albert2020-109065}.
An implicit transformation to canonical coordinates is also required to preserve
symplectic structure in variational integrators~\citep{Kraus2017, Burby2017-110703, Ellison2018-52502}. There, the transformation to a form where canonical momenta can be identified is referred to as a \emph{gauge} choice with an existence proof. Below, we describe a procedure to actually construct this transformation in practice.

Implicit spatial transformations to obtain canonical guiding-center
coordinates have been realized for magnetic flux coordinates to first
order in gyroradius~\citep{White1984-2455a,Boozer2005-1071,Boozer2015-025001}
and exactly~\citep{Zhang2014-32504,Li2016-334,Albert2020-109065}. Such coordinates
exist only in configurations with nested magnetic flux surfaces and
exclude, for example, the tracing of particles up to the wall of a
magnetic confinement configuration with a divertor. Also in the case of magnetic perturbations,
flux surface geometry is destroyed, and the topology of the magnetic
field includes islands and ergodic regions. \citep{Zhang2014-32504} also provide a canonicalization in general field topology as a series expansion that requires a transformation in both, guiding-center position and velocity space coordinates.

A general and straightforward construction of canonical coordinates by a purely spatial transformation has been proposed
by \citet{Meiss1990-2563}. This procedure consists of two steps:
a coordinate transformation to eliminate one covariant component of
the magnetic field, and a gauge transformation for the according covariant
vector potential component. In the present work, we build on that
approach of ``Meiss-Hazeltine coordinates'' with three adjustments for practical applicability:
1) The toroidal angle is modified instead of a poloidal variable.
This directly yields a unique-valued transformation in usual toroidal
magnetic confinement devices where the toroidal magnetic field component
is never zero. 2) Based on this variant, canonical coordinates of
cylindrical topology are constructed in addition to toroidal flux-like coordinates.
3) One vector potential component is used as the radial variable, thereby
simplifying analytical derivations and numerical implementations.
In particular, we show that the obtained coordinates are a generalization
of magnetic flux coordinates.

The described spatial coordinate transformation and resulting symplectic integrators are implemented
in the code SIMPLE~\citep{Albert2020-109065,Albert2020-815860201}.
Results from numerical experiments confirm the validity and efficiency
of symplectic orbit integration in both, tokamaks with broken symmetry and stellarators with magnetic islands and stochastic
regions.

\section{Transformation to canonical coordinates}

We write the guiding-center Lagrangian $L_{\mathrm{gc}}$ with Hamiltonian
$H$ skipping the term $J_{\perp}\dot{\phi}$ of the ignorable pair
as

\begin{align}
L_{\mathrm{gc}}(\v x,z^{4})= & \frac{e_{\alpha}}{c}\left(A_{1}^{\star}(\v x,z^{4})\dot{x}^{1}+A_{2}^{\star}(\v x,z^{4})\dot{x}^{2}\right.\nonumber \\
 & \left.+A_{3}^{\star}(\v x,z^{4})\dot{x}^{3}\right)-H(\v x,z^{4}) \label{eq:Lgc0}
\end{align}
in terms of non-canonical coordinates $\v z=(x^{1},x^{2},x^{3},z^{4})$,
where
\begin{equation}
A_{i}^{\star}(\v x,z^{4})=\frac{v_{\parallel}(\v x,z^{4})}{\omega_{c\alpha}(\v x)}B_{i}(\v x)+A_{i}(\v x) \label{eq:astar}
\end{equation}
contain both, covariant components $A_{i}$ of the vector potential
and $B_{i}$ of the magnetic field. In addition, parallel velocity
$v_{\parallel}$, speed of light $c$, species charge $e_{\alpha}$
and gyrofrequency $\omega_{c\alpha}$ appear. Choices of the fourth
phase-space coordinate $z^{4}$ besides three spatial variables $\v x$
include $v_{\parallel}$ itself, Hamiltonian $H$, or the toroidal
canonical momentum $p_{\varphi}$. As a first step~\citep{Meiss1990-2563,Albert2020-109065},
we transform to new spatial coordinates $x_{\text{c}}^{1},x_{\text{c}}^{2},x_{\text{c}}^{3}$
(subscript ``c'' for ``canonical'') with the requirement to eliminate the first covariant component of
$\v B$, so
\begin{align}
B_{1\mathrm{c}}(\v x_{\c})= & \frac{\partial x^{i}(\v x_{\c})}{\partial x_{\mathrm{c}}^{1}}B_{i}(\v x(\v x_{\c}))\overset{!}{=}0.
\end{align}
Assuming prescribed transformations for two components $x^{1}(\v x_{\c})$
and $x^{3}(\v x_{\c})$ this requires
\begin{equation}
\frac{\partial x^{2}(\v x_{\c})}{\partial x_{\mathrm{c}}^{1}}=-\frac{\frac{\partial x^{1}(\v x_{\c})}{\partial x_{\mathrm{c}}^{1}}B_{1}(\v x(\v x_{\c}))+\frac{\partial x^{3}(\v x_{\c})}{\partial x_{\mathrm{c}}^{1}}B_{3}(\v x(\v x_{\c}))}{B_{2}(\v x(\v x_{\c}))},\label{eq:dx3}
\end{equation}
under the restriction that $B_{2}$ never vanishes. For any fixed pair $x_{\c}^{2},x_{\c}^{3}$, Eq.~(\ref{eq:dx3}) is a nonlinear ordinary differential equation with independent variable
$x_{\c}^{1}$ .

To eliminate the component $A_1^\c(\v x_{\c})$  in Eq.~\eqref{eq:astar}, a gauge transform $\v A^\c = \v A+\nabla\chi$ is applied~\citep{Meiss1990-2563} in addition to the covariant coordinate transform.  Thus, condition $A_1^\c=0$  yields an ordinary differential equation for the gauge function $\chi(\v x_{\c})$,
\begin{align}
\frac{\partial\chi(\v x_{\c})}{\partial x_{\mathrm{c}}^{1}} & =-\frac{\partial x^{i}(\v x_{\c})}{\partial x_{\mathrm{c}}^{1}}A_{i}(\v x(\v x_{\c})).\label{eq:dA1c}
\end{align}
with $\partial x^{2}/\partial x_{\mathrm{c}}^{1}$ given by Eq.~(\ref{eq:dx3}) .
The solution of Eqs.~(\ref{eq:dx3}) and (\ref{eq:dA1c}) is found via
a generic integrator for ordinary differential equations. Thus, we
have eliminated $A_{1}^{\star \c}$ in Eq.~\eqref{eq:astar}, which allows us to identify canonical
momenta
\begin{equation}
p_{2}=\frac{e_\alpha}{c}A_{2}^{\star\c}(\v x_{\c},z^{4})\text{ and }p_{3}=\frac{e_\alpha}{c}A_{3}^{\star\c}(\v x_{\c},z^{4})\label{eq:momenta}
\end{equation}
paired to $x_{\mathrm{c}}^{2}$ and $x_{\mathrm{c}}^{3}$, respectively,
inside $L_{\mathrm{gc}}(\v x_{\c},z^{4})$ in Eq.~\eqref{eq:Lgc0}. Note that this formulation
still contains two generally non-canonical coordinates, $z^{1}=x_{\text{c}}^{1}$
and $z^{4}$. Now, values of these coordinates can be computed implicitly
from Eq.~\eqref{eq:momenta} for points given in canonical coordinates
$(x_{\text{c}}^{2},p_{2},x_{\text{c}}^{3},p_{3})$. This enables the application of symplectic~\citep{Albert2020-109065} and variational~\citep{Burby2017-110703,Ellison2018-52502} integrators to the guiding-center system.

Let us now consider special cases useful in toroidal magnetic configurations,
such as tokamaks or stellarators with non-zero toroidal component $B_{\ph}$ everywhere in the domain of interest. We fix $x^2=\varphi$ to a toroidal angle and choose a transformation with

\begin{align}
x^{1} & =x_{\c}^{1},\nonumber \\
\varphi & =\varphi_{\c}+\lambda(x_{\c}^{1},\varphi_{\c},x_{\c}^{3}),\nonumber \\
x^{3} & =x_{\c}^{3}.
\end{align}
In cylindrical-type coordinates, we have $x^{1}=R,\,x^{3}=Z$, and
in flux-like coordinates, $x^{1}=r$ is a radial coordinate and $x^{3}=\vartheta$
a poloidal angle. Eqs.~(\ref{eq:dx3}-\ref{eq:dA1c}) become
\begin{equation}
\frac{\partial\lambda}{\partial x_{\c}^{1}}=-\frac{B_{1}}{B_{\varphi}},\quad\frac{\partial\chi}{\partial x_{\c}^{1}}=-A_{1}+\frac{B_{1}}{B_{\varphi}}A_{\varphi}.\label{eq:dx3-1}
\end{equation}
In an additional step, we replace $x_{\c}^{1}$ by using the vector
potential component

\begin{equation}
A_{3}^{\c}=A_{3}+\frac{\partial\lambda}{\partial x_{\c}^{3}}A_{\varphi}+\frac{\partial\chi}{\partial x_{\c}^{3}}\equiv\psi_{\text{c}}
\end{equation}
to replace the first coordinate $x_{\text{c}}^{1}$ ($A_3^\c$ is a monotonous function of the original $x^1_\c$ due to non-vanishing contra-variant toroidal field component $B^\varphi_c$ as follows from $\partial A_3^\c/\partial x^1_\c = -\sqrt{g_c} B^\varphi_c$). This step does
not affect values of covariant vector field components. Similar to
a magnetic flux label, using $\psi_{\text{c}}$ as a coordinate has
the advantage that $A_{3}^{\c}$ will depend (identically) on $\psi_{\text{c}}$
only. Following the convention of magnetic flux coordinates, we finally
rename $x_{\c}^{3}\equiv\vartheta_{\c}$. In these final spatial coordinates
$\v x_\c = (\psi_{\text{c}},\varphi_{\c},\vartheta_{\c})$ we express the metric
determinant by magnetic quantities via
\begin{equation}\label{eq:MetricDeterminant}
\sqrt{g_{\c}}=\frac{1}{B^{2}}\left(B_{\vartheta}^{\c}\difp{A_{\varphi}^{\c}}{\psi_{\c}}-B_{\varphi}^{\c}\right).
\end{equation}
Combined with the choice $z^{4}\equiv A_{\varphi}^{\star\c} = p_{\varphi}$,
the guiding-center Lagrangian takes the form
\begin{equation}
L_{\mathrm{gc}}(\v x_{\c},p_{\varphi}) =p_{\varphi}\dot{\varphi}_{\text{c}}+p_{\vartheta}(\v x_{\c},p_{\varphi})\dot{\vartheta}-H(\v x_{\c},p_{\varphi}).\label{eq:Lgc1}
\end{equation}
Here, poloidal canonical momentum and parallel velocity respectively are
\begin{align}
p_{\vartheta}(\v x_{\c},p_{\varphi}) & =m_{\alpha}v_{\parallel}(\v x_{\c},p_{\varphi})\frac{B_{\vartheta}^{\c}(\v x_{\c})}{B(\v x_{\c})}+\frac{e_{\alpha}}{c}\psi_{\text{c}},\label{eq:pth}\\
v_{\parallel}(\v x_{\c},p_{\varphi}) & =\frac{B(\v x_{\c})}{m_\alpha B^{\c}_{\varphi}(\v x_{\c})}\left(p_{\varphi}-\frac{e_\alpha}{c}A_{\varphi}^{\c}(\v x_{\c})\right),
\end{align}
and the Hamiltonian is
\begin{equation}
H(\v x_{\c},p_{\varphi}) =\frac{m_{\alpha}v_{\parallel}^{\,2}(\v x_{\c},p_{\varphi})}{2}+J_{\perp}\omega_{c\alpha}(\v x_{\c})+e_{\alpha}\Phi_{e}(\v x_{\c}),\label{eq:Lgc2}
\end{equation}
with perpendicular invariant $J_{\perp}$ and electric potential
$\Phi_{e}$. Magnetic field line equations are recovered from the Euler-Lagrange equations obtained from Eqs.~(\ref{eq:Lgc1}-\ref{eq:Lgc2}) with $J_\perp=\Phi_e=0$ using $z^4=v_\parallel$ instead of $p_\varphi$. 
Only two out of three equations corresponding to spatial variables are independent in the limit $m_{\alpha}/e_{\alpha}\rightarrow0$,
\begin{equation}
 \dot \psi_\c = \frac{\partial A_\varphi^\c}{\partial \vartheta_\c} \dot\varphi_\c,
 \qquad
 \dot \vartheta_\c = - \frac{\partial A_\varphi^\c}{\partial \psi_\c} \dot\varphi_\c, \label{eq:zeroflr}
\end{equation}
and the fourth equation is $B_\vartheta^\c \dot\vartheta_\c + B_\varphi^\c \dot\varphi_\c = B v_\parallel$, so that $\dot x^k_\c = v_\parallel B^k_\c /B$. Replacing time with a new independent variable $\varphi_\c$, Eqs.~(\ref{eq:zeroflr}) turn into field line equations in the canonical Hamiltonian form for a gauge 
 with $A_{1}^{\c} = 0$ (see, e.g.,~\citep{Abdullaev2006-S113,Boozer2015-025001}).
In addition, our toroidal angle $\varphi_{\text{c}}$ is different
from the cylinder angle $\varphi$, which is not necessary for the
canonical representation of field lines, but essential for guiding-centers
where we require $B_{1}^{\c}=0$ via Eq.~(\ref{eq:dx3}).

\section{Symplectic orbit integration}

Symplectic integration schemes with an implicit transformation to canonical coordinates~\citep{Zhang2014-32504, Albert2020-109065} have been implemented
on the new canonical coordinates in the code SIMPLE. Here, the implicit transformation
from canonical coordinates~\eqref{eq:momenta} is solved together with the canonical symplectic time-stepping scheme for the Hamiltonian flow. For the guiding-center system in the present coordinates $\v z = (\psi_\c,\varphi_\c,\vartheta_\c,p_\varphi)$, the Hamiltonian flow for the according canonical coordinates $\v z_\c = (p_\vartheta,\varphi_\c,\vartheta_\c,p_\varphi)$ is given by canonical phase-space velocity components $V_\c^k$ as functions of non-canonical $\v z$ (Eqs.~(63-66) of~\cite{Albert2020-109065}),
\begin{align}
    V_\c^{p_{\vartheta}} &= -\frac{\partial H}{\partial\tht}+\frac{\partial H}{\partial \psi_\c}\left(\frac{\partial p_{\tht}}{\partial \psi_\c}\right)^{-1}\frac{\partial p_{\tht}}{\partial\tht},\nonumber\\
    V_\c^\varphi &= \frac{B}{B_{\ph}^\c}\left(v_{\parallel}-\frac{\partial H}{\partial \psi_\c}\left(\frac{\partial p_{\tht}}{\partial \psi_\c}\right)^{-1}\frac{B_{\tht}^\c}{B}\right), \nonumber\\
    V_\c^\vartheta &= \frac{\partial H}{\partial \psi_\c}\left(\frac{\partial p_{\tht}}{\partial \psi_\c}\right)^{-1}, \nonumber\\
    V_\c^{p_{\varphi}} &= -\frac{\partial H}{\partial\ph}+\frac{\partial H}{\partial \psi_\c}\left(\frac{\partial p_{\tht}}{\partial \psi_\c}\right)^{-1}\frac{\partial p_{\tht}}{\partial\ph}
    .\label{eq:Vphase}
\end{align}
For the symplectic midpoint rule~\cite{Hairer2006, Zhang2014-32504} with timestep $h$, implicit equations in non-canonical $\v z_{(n+1/2)}$ that represent the midpoint $\frac{1}{2}(\v z_{\c,(n)} + \v z_{\c,(n+1)})$ in canonical coordinates are
\begin{align}
    p_{\vartheta}(\v z_{(n+1/2)}) &= p_{\vartheta,(n)} + \frac{\Delta t}{2} V_\c^{p_{\vartheta}}\left(\v z_{(n+1/2)} \right), \label{eq:forpt}\\
    z^k_{(n+1/2)} &= z^k_{\c,(n)} + \frac{\Delta t}{2} V_\c^k\left(\v z_{(n+1/2)}\right), \quad k > 1,    
\end{align}
where the left-hand side of Eq.~(\ref{eq:forpt}) is computed from Eq.~\eqref{eq:pth}. The remaining equations for the next step \mbox{$(n+1)$} become explicit with
\begin{equation}
    z^k_{\c,(n+1)} = z^k_{\c,(n)} + \Delta t\,V_\c^k\left(\v z_{(n+1/2)}\right), 
\end{equation}
re-using values of the phase-space velocity~\eqref{eq:Vphase} evaluated at the midpoint $\v z_{(n+1/2)}$.

Orbits are traced in the magnetic field of a typical medium-sized tokamak configuration with a lower single-null divertor and resonant magnetic perturbations as in Refs~\citep{martitsch2016,suttrop2018}. We compare the symplectic scheme to two non-symplectic adaptive methods: Runge-Kutta-Fehlberg of order 4/5 and Dormand\&Prince of order 8/5/3~\citep{Hairer1993,Dormand2018}. Each setup has been chosen such that the total number of magnetic field evaluations is similar for all compared integration methods.

\begin{figure}
\includegraphics{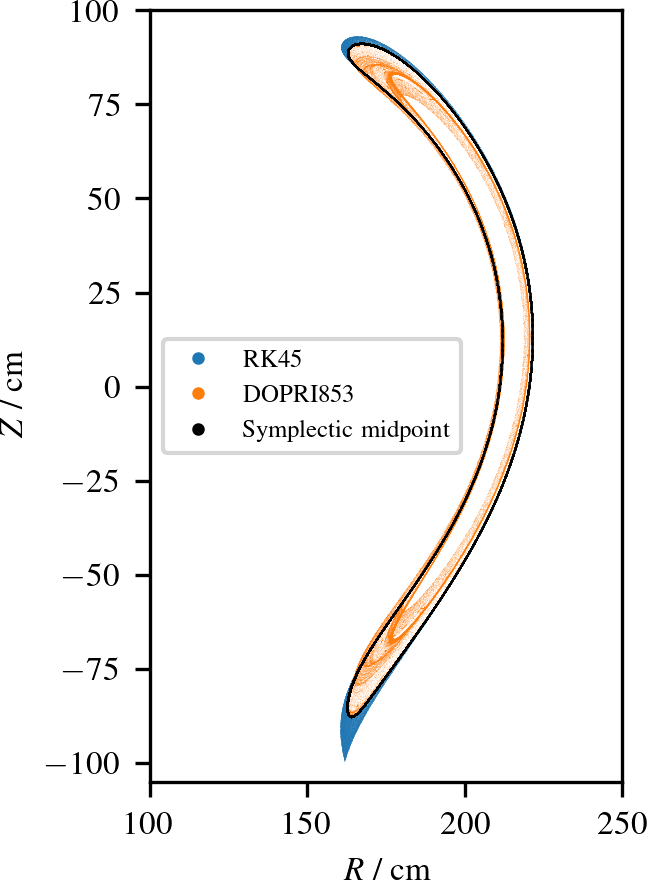}

\caption{Banana orbit crossing an axisymmetric tokamak's separatrix. Results
from symplectic midpoint (black)~\cite{Hairer2006, Albert2020-109065} and non-symplectic RK45 (blue) and DOPRI853 (orange)~\cite{Hairer1993,Dormand2018}. Orbits from non-symplectic
methods shrink or escape after a few 10000 bounces in contrast to symplectic
ones at similar computational cost in terms of field evaluations. \label{fig:banana}}
\end{figure}

\begin{figure}
(a) \includegraphics{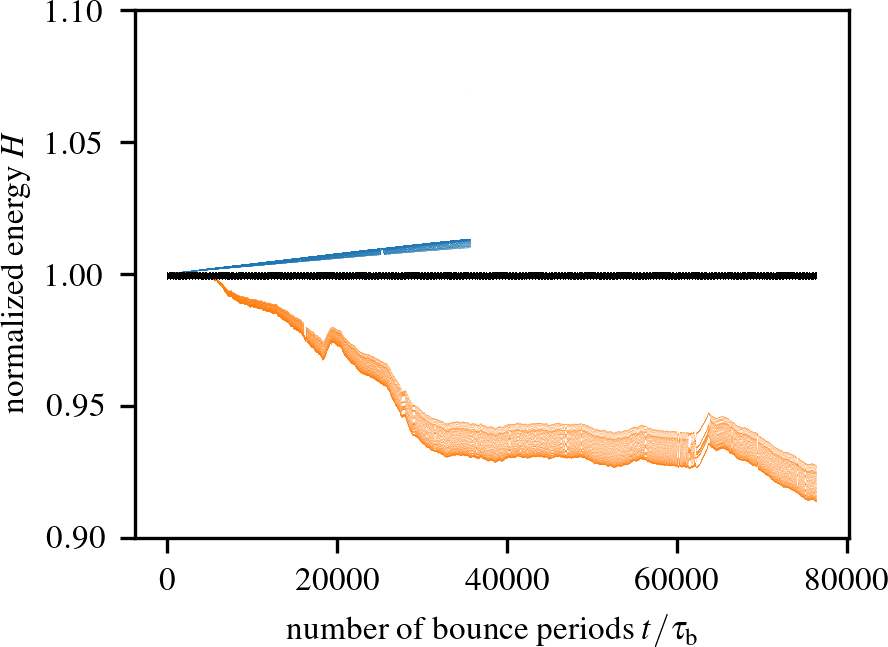}

(b) \includegraphics{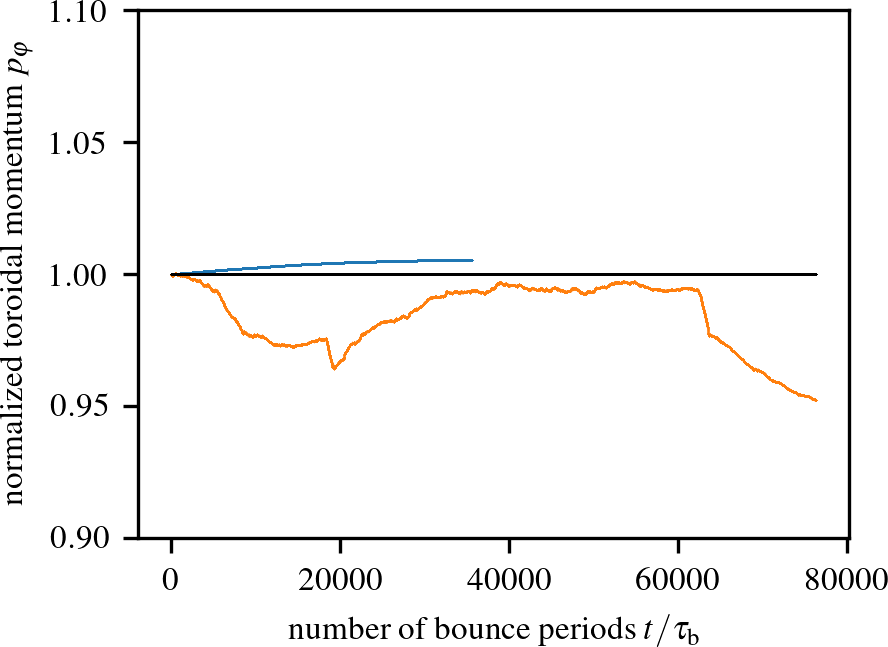}

\caption{Energy $H$ (a) and toroidal momentum $p_{\varphi}$ (b) of the orbit
of Fig. 1 (same coloring) over time $t$ normalized by their initial value. Values drift away for
non-symplectic methods RK45 and DOPRI853 and are truncated when the orbit incorrectly leaves the device for RK45. The implicit midpoint rule ensures exact conservation
($p_{\varphi}$) or oscillation in bounds ($H$) due to symplecticity. \label{fig:energy}}
\end{figure}

Fig.~\ref{fig:banana} shows the orbit of a trapped "banana" orbit in the axisymmetric field, with energy and toroidal momentum evolution plotted in Fig.~\ref{fig:energy}. While the symplectic midpoint integrator retains the shape and integrals of motion, the non-symplectic methods deviate up to a destruction of the orbit topology.

\begin{figure}
\includegraphics{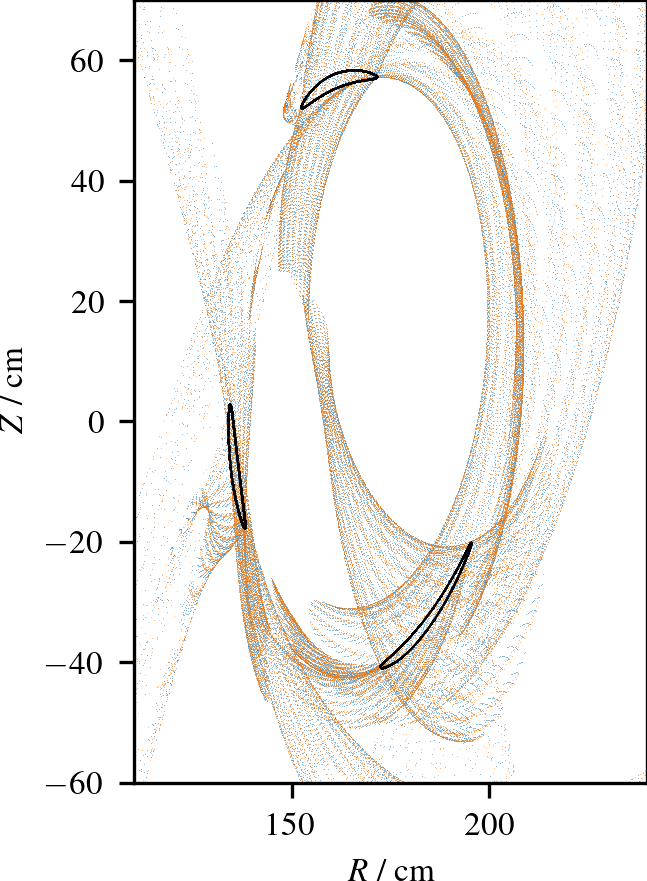}

\caption{Island chain formed by Poincaré sections of passing orbits at constant toroidal angle in a perturbed tokamak field. Symplectic midpoint (black) preserves its structure, opposed to RK45 (blue) and DOPRI853 (orange). \label{fig:islands}}
\end{figure}

\begin{figure}
\includegraphics{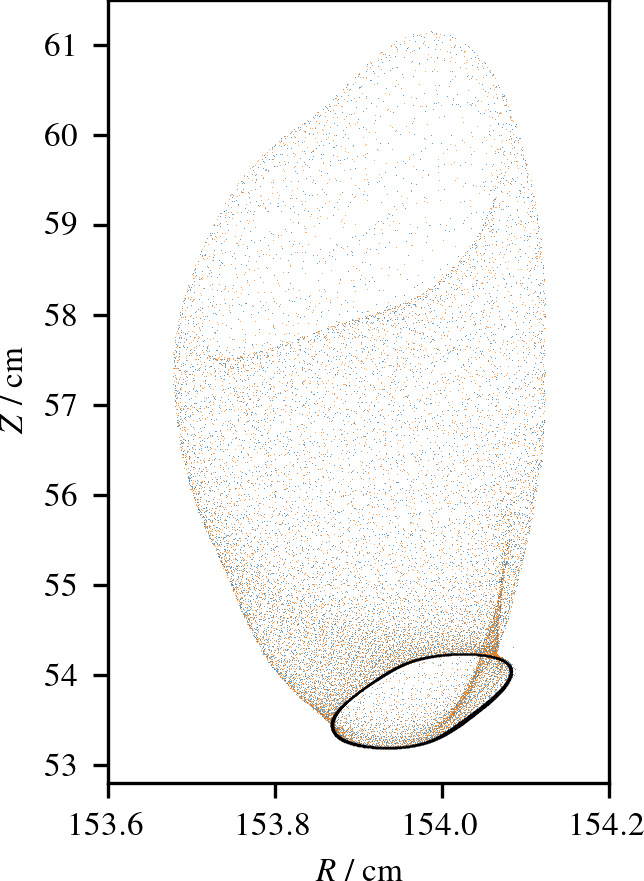}

\caption{Banana tip precession by Poincaré sections of trapped orbits at zero $v_\parallel$ in a perturbed tokamak field. Symplectic midpoint (black) preserves its structure, opposed to RK45 (blue) and DOPRI853 (orange). \label{fig:bananatip}}
\end{figure}

Figs.~\ref{fig:islands} and~\ref{fig:bananatip} show the behavior of passing and trapped orbits in a perturbed tokamak field. We have chosen a region where non-linear oscillations appear around a phase-space resonance. The symplectic integrator preserves the structure of the islands and the precession of the banana tip, while the non-symplectic methods fail to do so. This is especially relevant for transport induced by such resonances, that can be shown to be a main mechanism for the loss of alpha particles~\citep{Paul2022-126054} as well as toroidal viscosity~\citep{Park2009-065002} in 3D fields. Both features are critical for stable and efficient operation of magnetic confinement fusion devices.


\section{Conclusion and Outlook}

Symplectic integration of the guiding-center system of charged particle dynamics
in strong magnetic fields has been realized via an effective transformation
to canonical coordinates. Starting with a short review of previously
available methods, we derived a variant of Meiss-Hazeltine coordinates for this purpose.
In contrast to related approaches, this transformation requires neither
nested magnetic flux surfaces nor a full phase-space transformation.
Starting with the general description, we pursue a realization for cylindrical
coordinates. Based on these canonical coordinates, we implemented
the symplectic midpoint rule based on the SIMPLE algorithm. For realistic
conditions in a tokamak field with resonant magnetic perturbations,
the method outperforms conventional non-symplectic integrators to
a similar degree as observed in previous works. This is significant
for efficient and physics-consistent numerical implementations of
drift- and gyrokinetic models from the plasma core to the wall. Due
to their similarity to canonical magnetic flux coordinates,
the new canonical coordinates are also of interest to generalize
theoretical results that previously relied on nested magnetic flux
surfaces.
\newpage
\begin{acknowledgments}
The authors thank Benedikt Brantner, Martin Heyn, Michael Kraus and Pedro Gil for useful discussions. This work has been carried out within the framework of the EUROfusion Consortium, funded by the European Union via the Euratom Research and Training Programme (Grant Agreement No 101052200 — EUROfusion). Views and opinions expressed are however those of the author(s) only and do not necessarily reflect those of the European Union or the European Commission. Neither the European Union nor the European Commission can be held responsible for them. We gratefully acknowledge support from NAWI Graz.
\end{acknowledgments}

\bibliography{canonical}

\end{document}